\documentclass{aa}
\usepackage{txfonts}
\usepackage{graphicx}
\usepackage{threeparttable}
\usepackage{natbib}

\begin{document}

\title{Resolving the inner regions of the HD97048 circumstellar disk with VLT/NACO Polarimetric Differential Imaging\thanks{Based on observations collected at the European Organisation for Astronomical Research in the Southern Hemisphere, Chile (program number: 077.C-0106A).}}

\author{Sascha P. Quanz\inst{1} 
\and Stephan M. Birkmann\inst{2} 
\and Daniel Apai\inst{3} 
\and Sebastian Wolf\inst{4} 
\and Thomas Henning\inst{5} 
}

\institute{Institute for Astronomy, ETH Zurich, Wolfgang-Pauli-Strasse 27, CH-8093 Zurich, Switzerland
\and ESA/ESTEC, Keplerlaan 1, Postbus 299, 2200 AG Noordwijk, The Netherlands
\and Department of Astronomy/Department of Planetary Sciences, University of Arizona, 933 N. Cherry Ave., Tucson, AZ 85721, USA
\and University of Kiel, Institute of Theoretical Physics and Astrophysics, Leibnizstrasse 15, 24098 Kiel, Germany
\and Max Planck Institute for Astronomy, K\"onigstuhl 17, 69117 Heidelberg, Germany}


\abstract{Circumstellar disks are the cradles of planetary systems and their physical and chemical properties directly influence the planet formation process. As most planets supposedly form in the inner disk regions, i.e., within a few tens of AU, it is crucial to study circumstellar disk on these scales to constrain the conditions for planet formation.}
{Our aims are to characterize the inner regions of the circumstellar disk around the young Herbig Ae/Be star HD97048 in polarized light.}{We use VLT/NACO to observe HD97048 in polarimetric differential imaging (PDI) mode in the $H$ and $K_s$ band. PDI offers high-contrast capabilities at very small inner working angles and probes the dust grains on the surface layer of the disk that act as the scattering surface.}{We spatially resolve the disk around HD97048 in polarized flux in both filters on scales between $\sim$0.1$''$--1.0$''$ corresponding to the inner $\sim$16--160 AU. Fitting isophots to the flux calibrated $H$-band image between 13 -- 14 mag/arcsec$^2$ and 14 -- 15 mag/arcsec$^2$ we derive a apparent disk inclination angle of 34$^\circ\pm5^\circ$ and 47$^\circ\pm2^\circ$, respectively. The disk position angle in both brightness regimes is almost identical and roughly 80$^\circ$. Along the disk major axis the surface brightness of the polarized flux drops from $\sim$11 mag/arcsec$^2$ at $\sim$0.1$''$ ($\sim$16 AU) to $\sim$15.3 mag/arcsec$^2$ at $\sim$1.0$''$ ($\sim$160 AU). The brightness profiles along the major axis are fitted with power-laws falling off as $\propto$$r^{-1.78\pm0.02}$ in $H$ and $\propto$$r^{-2.34\pm0.04}$ in $K_s$. As the surface brightness drops off more rapidly in $K_s$ compared to $H$ the disks becomes relatively bluer at larger separations possibly indicating changing dust grain properties as a function of radius.}{For the first time the inner $\sim$0.1$''$--1.0$''$ ($\sim$16--160 AU) of the surface layer of the HD97048 circumstellar disk have been imaged in scattered light demonstrating the power of ground-based imaging polarimetry. Our data fill an important gap in a large collection of existing data including resolved thermal dust and PAH emission images as well as resolved gas emission lines. HD97048 is thus an ideal test case for sophisticated models of circumstellar disks and a prime target for future high-contrast imaging observations.}

\titlerunning{NACO/PDI observations of the HD97048 circumstellar disk}
\authorrunning{Quanz et al.}

\maketitle

\section{Introduction}
Planets form in disks of gas and dust around young stars and models show that the physical and chemical properties of these disks influence the outcome of the planet formation process \citep[e.g.,][]{alibert2011}. At least for sun-like stars most direct imaging surveys for exoplanets found that (massive) planets on large orbital separations ($\gtrsim$ 50 AU) are rather scarce \citep[e.g.,][]{masciadri2005,biller2007,kasper2007,apai2008,lafreniere2007,chauvin2010,janson2011} but radial velocity and transit searches have demonstrated that the inner few AU are packed with (low-mass) planets \citep[e.g.,][]{mayor2011,borucki2011}. The direct detection of giant planets orbiting between $\sim$10--70 AU around the A-type stars $\beta$ Pictoris \citep{lagrange2009a,lagrange2010,quanz2010} and HR8799 \citep{marois2008,marois2010} suggest that at least intermediate mass stars can form planets out to a few tens of AU. Thus, in order to study those regions in circumstellar disks where planets are potentially forming one requires an observing technique capable of resolving those inner regions of circumstellar disks.

Polarimetric Differential Imaging (PDI) is a technique that allows high-contrast direct imaging of circumstellar disks at very small inner working angles \citep{kuhn2001,potter2003,apai2004,perrin2004}. PDI takes advantage of the fact that direct stellar light is largely unpolarized while stellar photons that scatter of the dust grains on the surface layer of a circumstellar disk are polarized. In a recent paper we presented near-infrared (NIR) PDI observations of the circumstellar disk around the Herbig Ae/Be star 100546 \citep{quanz2011}. These observations revealed disk structures, geometry and dust grain properties as close as 0.1$''$ to the central star, i.e., regions that are difficult to probe by other direct imaging techniques in the NIR such as coronagraphy or classical PSF subtraction.

In this paper we present PDI data of HD97048 another well-studied Herbig Ae/Be star. The key stellar parameters are summarized in Table~\ref{object}. We note that throughout this paper we assume a distance of 158$^{+16}_{-14}$ pc for HD97048 as derived from a re-analysis of the Hipparcos measurements by \citet{vanleeuwen2007}. Most earlier studies used the initial Hipparcos results of 180$^{+30}_{-20}$ pc found by \citet{vandenancker1997}\footnote{Already \citet{whittet1997} favored a distance of only 160 pc  consistent with the distance to the Chameleon I region. We emphasize, however, that the error bars for both derived distances overlap and that the only figure in this paper that directly depends on the exact value for the distance is the projected separation of circumstellar material around the star}.

HD97048 is surrounded by a circumstellar disk that was clearly resolved in mid-infrared (MIR) images taken in PAH filters \citep{lagage2006,doucet2007}. Spectroscopic studies in the MIR showed also  very strong and resolved PAH emission bands but no silicate emission feature at 10 $\mu$m \citep{vankerckhoven2002,vanboekel2004}. \citet{doering2007} imaged the circumstellar surrounding of HD97048 with HST/ACS in the F606W (broad V) filter and detected scattered light out to a radial distance of $\sim$4$''$ in almost all directions. The inner 2$''$ (in radius) were excluded from any analysis due to the occulting spot used for the observations and substantial subtraction residuals. Single-dish observations of the mm continuum emission suggest a circumstellar mass of $\sim$0.2 M$_{\sun}$ \citep{henning1998} but part of this mass is probably residing in an envelope, also seen in extended MIR emission \citep{prusti1994,siebenmorgen2000}, surrounding the star-disk system. Several gas emission lines ([OI], CO and H$_2$) have been detected and were, at least partly, spectrally and spatially resolved \citep{acke2006,vanderplas2009,martinzaidi2007,martinzaidi2009,carmona2011}. 

Our polarimetric images resolve the inner parts of the HD97048 circumstellar disk for the first time in scattered light in the $H$ and $K_s$ band. The observations and data reduction are described in section 2, the key results and main analyses are presented in section 3, and we discuss our results in broader context in section 4. Finally, we conclude in section 5. 


\begin{table} 
\caption{Properties of central star.} 
\label{object}	
\centering 
\begin{tabular}{llc} 
\hline\hline
 {Parameter} &  {HD97048} & {Reference}\\\hline
RA (J2000) & 11$^h$08$^m$03.32$^s$  & (1) \\
DEC (J2000) & -77$^\circ$39$'$17.48$''$ & (1) \\
H [mag] & 6.67  & (2) \\
Ks [mag] & 5.94 & (2) \\
Distance [pc] & 158$^{+16}_{-14}$ & (1) \\
Sp. Type &  B9-A0ep+sh & (3)\\
Age [Myr] & $>$2, 3   & (4),(5) \\
log (T$_{eff}$) [K] & 4.0 & (4) \\
A$_V$ [mag] & 1.24 & (4) \\
Mass [M$_{\sun}$] & 2.5$\pm$0.2 & (4)\\\hline
\end{tabular} 
\tablebib{
(1) \citet{vanleeuwen2007}; (2) \citet{cutri2003}; (3) \citet{whittet1987}; (4) \citet{vandenancker1998}; (5) \citet{lagage2006}}
\end{table}

\section{Observations and data reduction}
The observations were carried out with VLT/NACO \citep{lenzen2003,rousset2003} on ESO's UT4 in April 2006 and followed the same strategy as those used in \citet{quanz2011}. 
The SL27 camera with a pixel scale of $\sim$27 mas/pixel was used with the detector set to {\tt HighDynamic} mode and read out in {\tt Double RdRstRd} mode. For our polarimetric observations the Wollaston prism was used splitting the light into the ordinary and extraordinary beam offset by 3.5$''$ in $y$-direction
on the detector. The polarimetric mask avoids overlap of the two images on the detector but restricts the original field of view of ~27$''\times$27$''$ to $x=27''$ long and $y=3.1''$ wide stripes separated by 3.5$''$ in $y$-direction. To obtain full polarization cycles we used the rotatable half-wave retarder plate at four different positions ($0.0^\circ,-22.5^\circ,-45.0^\circ,-67.5^\circ$). The object was moved along the detector's x-axis between consecutive polarization cycles to correct for bad pixels. 

\begin{table*} 
\caption{Summary of observations and observing conditions.} 
\label{observations}	
\centering \begin{tabular}{cccccccc} \hline\hline
 {Object} &  {Filter} &  {DIT $\times$ NDIT$^{a}$} &  {Dither}  &   {Airmass} &  {Obs. date} &  {$\langle{EC}\rangle^{c}$[\%]} &  {$\langle \tau_0 \rangle^{d}$ [ms]}\\
 {} &  {} &  {} &  {positions$^{b}$}  &   {} &  {} &  {mean/min/max} &  {mean/min/max}\\\hline
\\
 HD97048 & $H$ & 0.35 s $\times$ 85 & 10(12) & 1.76 -- 1.68 & April 8, 2006 & 42.3/14.9/56.7 & 5.9/3.7/9.4\\
                   & $K_s$ & 0.35 s $\times$ 85 & 8(8)  & 1.68 -- 1.66 & April 8, 2006 & 50.2/41.3/55.9 & 5.5/2.7/8.1\\
 		& $NB1.64$ & 3.0 s $\times$ 20 & 16(16) & 1.66 -- 1.74 & April 8, 2006 & 53.9/46.2/60.4 & 7.3/3.0/9.8\\\hline
\end{tabular} 
\tablefoot{$^a$ Detector integration time (DIT) $\times$ number of integrations (NDIT), i.e., total integration time per dither position and per retarder plate position; $^b$ Number of dither positions used in final analysis and in parenthesis total number of observed dither position. The difference was disregarded due to poor AO correction. At each dither position NDIT exposures were taken at each of the 4 different retarder plate positions (0.0$^\circ$, -22.5$^\circ$, -45.0$^\circ$, -67.5$^\circ$); $^c$ Average, minimum and maximum value of the coherent energy of the PSF. Calculated by the Real Time Computer of the AO system; $^d$ Average, minimum and maximum value of the coherence time of the atmosphere. Calculated by the Real Time Computer of the AO system.}
\end{table*}

We observed HD97048 in three different filters ($H$, $K_s$, $NB1.64$). The core of the PSF was saturated in the $H$ and $K_s$ filter so that the inner 5-6 pixels (in diameter) were no longer in the linear detector regime (i.e. $>$10000 counts). The vast majority of the frames taken in the narrow band filter were unsaturated and only individual frames showed count rates slightly above the linearity threshold. The FWHM of the PSF in the $NB1.64$ was typically 4.0-4.5 pixels (i.e., 0.11$''$--0.12$''$) wide. Table~\ref{observations} summarizes the observations and the observing conditions.

The data reduction was laid out in \citet{quanz2011} and we refer the reader to this paper for a very detailed description. In short, after basic image processing all images were aligned with sub-pixel accuracy and for each dither position we computed the fractional polarization images $p_Q$ and $p_U$ in both filters. Averaging over all dither positions gave us the final images $\overline{p_Q}$ and $\overline{p_U}$. To obtain the final Stokes $Q$ and $U$ images  we computed the average intensity images $I$ for each filter and multiplied $Q=\overline{p_Q}\cdot I$ and $U=\overline{p_U}\cdot I$. The total polarized flux images $P$ were then derived via $P=(Q^2+U^2)^\frac{1}{2}$.

Also identical to the approach explained in \citet{quanz2011} the instrumental polarization was determined and corrected for before the final fractional polarization images were computed. We found the following offsets: 0.01102 and -0.00770 in the $H$ band for $p_Q$ and $p_U$, respectively, and 0.00018 and -0.00193 for the corresponding parameters in the $K_s$ filter. Finally, since we only obtained unsaturated images in the $NB1.64$ filter we could only flux calibrate the final $P$ image in the $H$ filter \citep[see,][]{quanz2011}. We estimate that the photometric calibration of the final $H$ band image is good to 40-50\% given the uncertainties in the instrumental polarization and in the calibration itself.

\begin{figure}
\includegraphics[width=4.45cm]{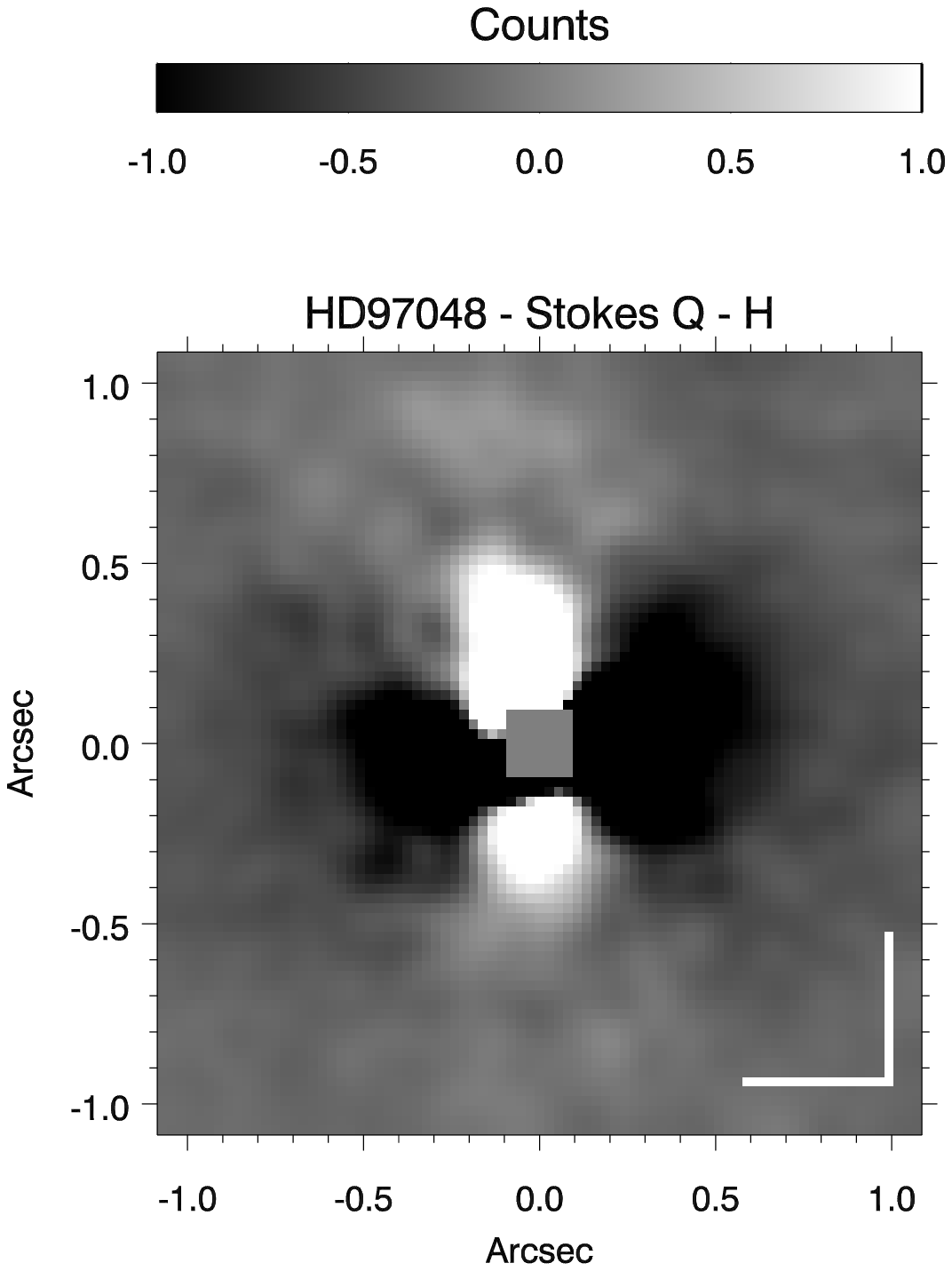}
\includegraphics[width=4.45cm]{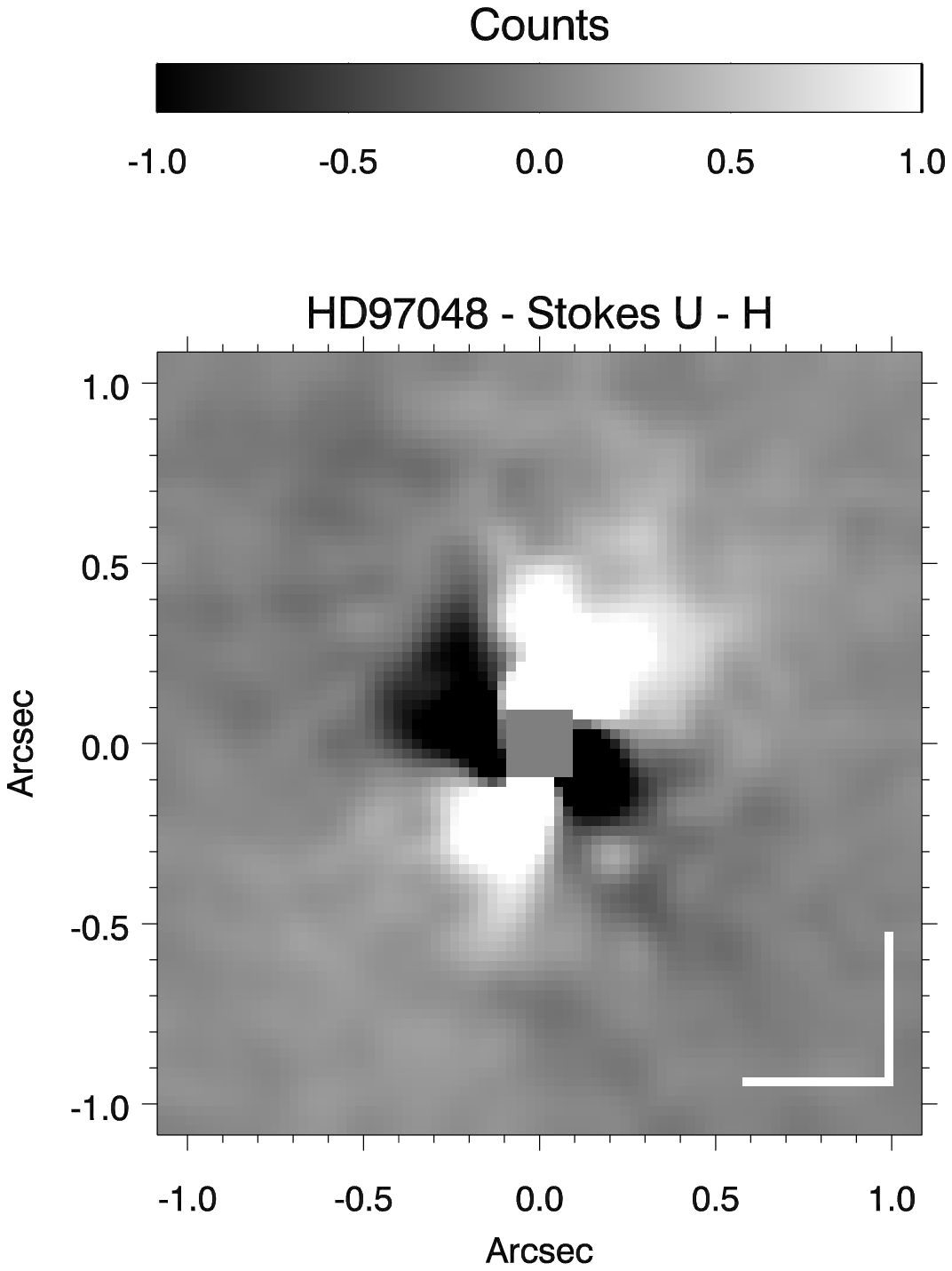}
\caption{Final Stokes Q (left) and Stokes U (right) images of HD97048 in the H filter. The innermost saturated pixels have been masked out and the images have been convolved with a Gaussian kernel with a FWHM of 4 pixels (i.e., the  FWHM of the PSF). The expected 'butterfly' pattern is clearly detected. The units are given in counts per pixel. North is up, east to the left.}
\label{Q_and_U_image}
\end{figure}

\section{Results and analysis}
In Figure~\ref{Q_and_U_image} we show the final Stokes $Q$ and $U$ images of HD97048 in the $H$ filter. The expected "butterfly" pattern of a circumstellar disk observed in PDI  is apparent with the pattern being rotated by $\sim$45$^\circ$ between $Q$ and $U$. Stokes $Q$ is normally defined to be positive in the north-south direction but as explained in \citet{witzel2011} and \citet{quanz2011} there was an offset of $\sim$13.2$^\circ$ in the encoder position of NACO's half-wave plate compared to the actual position on the sky. Thus, the pattern in Figure~\ref{Q_and_U_image} is rotated by this amount counterclockwise. 

\begin{figure}
\includegraphics[width=4.45cm]{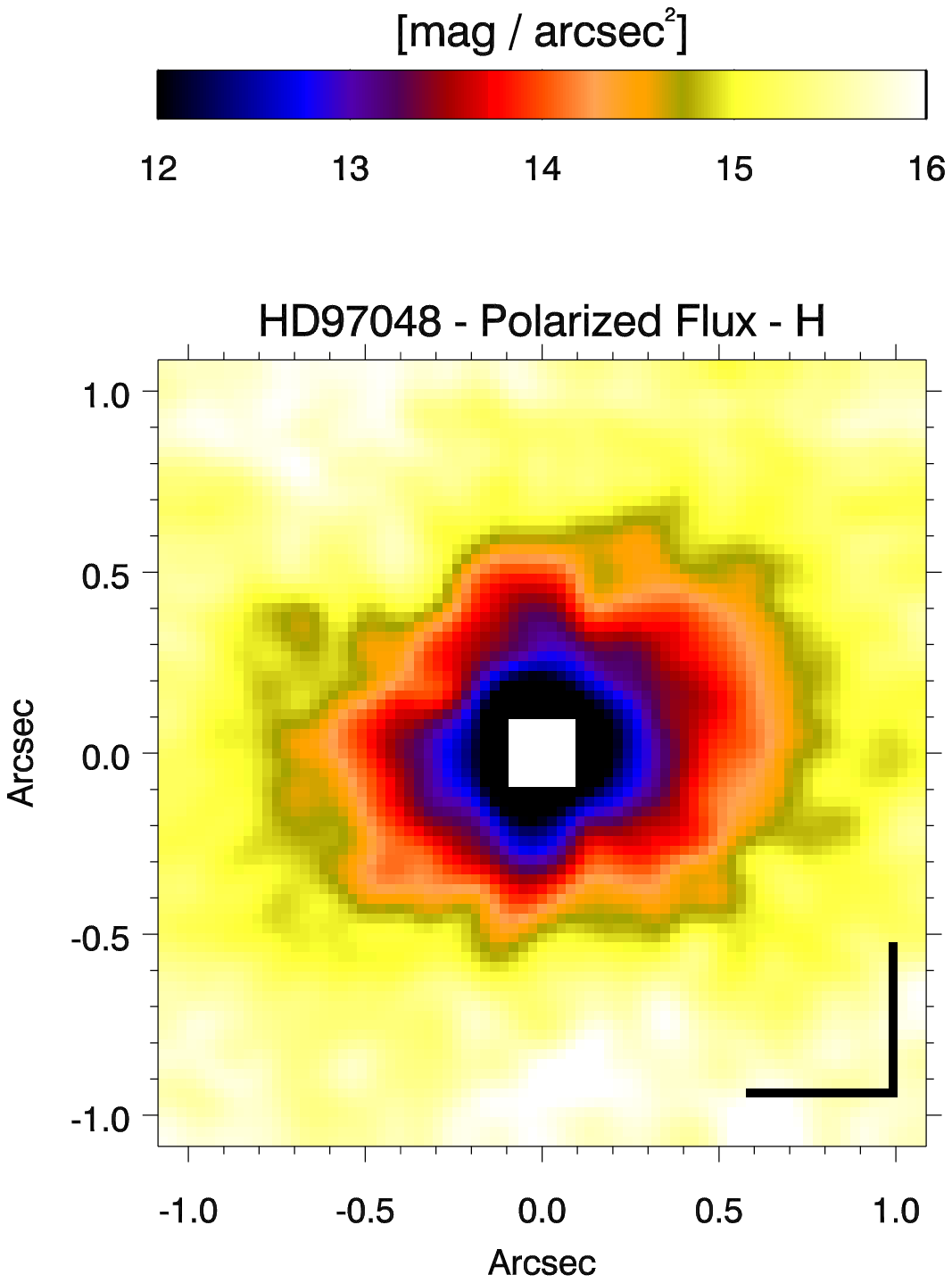}
\includegraphics[width=4.45cm]{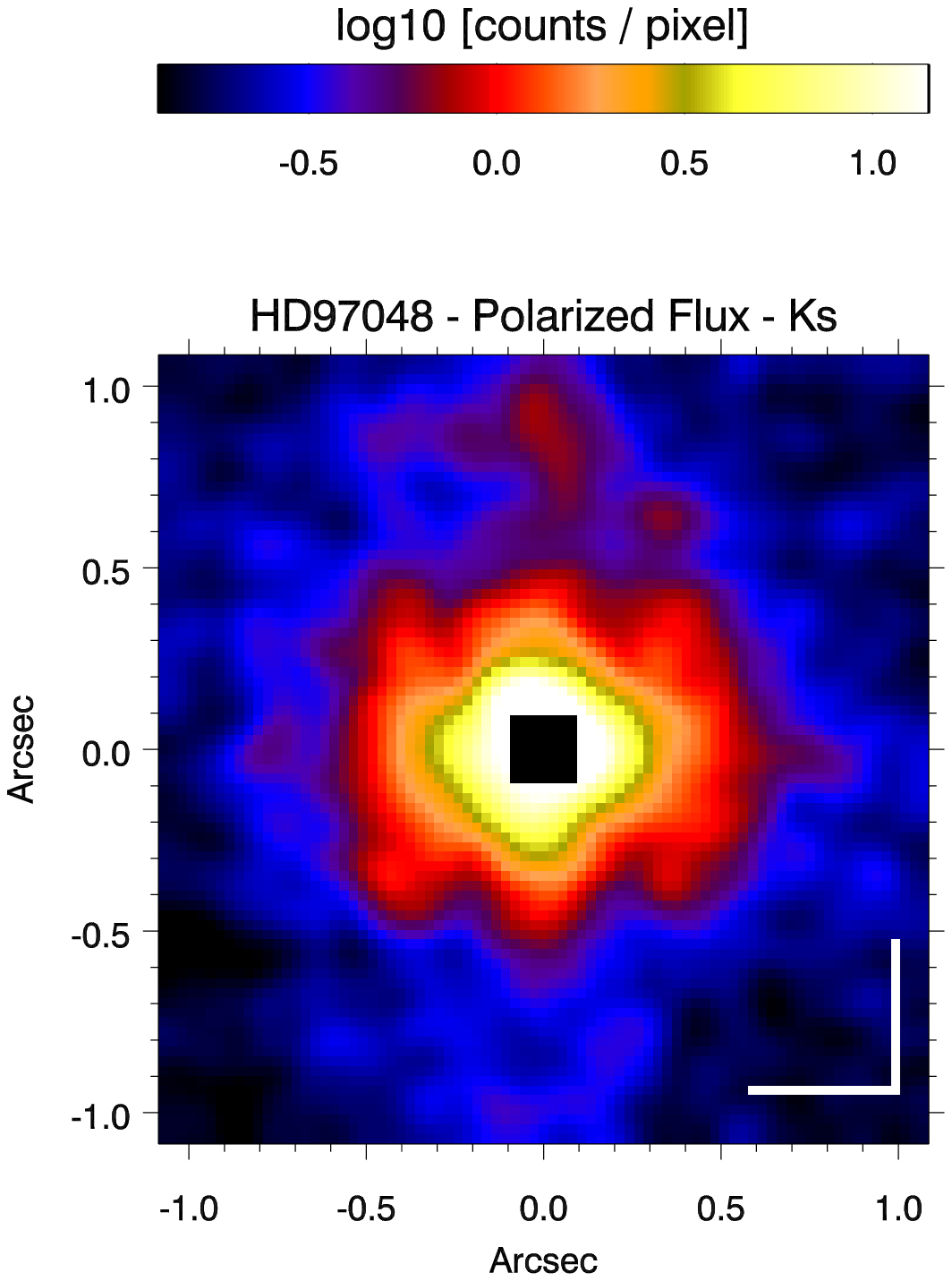}
\caption{Images showing the total polarized flux in the H filter (left) and Ks filter (right). The innermost saturated pixels have been masked out and the images have been convolved with a Gaussian kernel with a FWHM of 4 pixels (i.e., the FWHM of the PSF). While the H image is flux calibrated showing the flux in units of surface brightness, the unit of the Ks image is counts per pixel. North is up, east to the left.}
\label{P_images}
\end{figure}

In Figure~\ref{P_images} we show the final total polarized flux images in the $H$ and $K_s$ band. The $H$ band image is flux calibrated and the surface brightness of the disk is given in mag/arcsec$^2$. As the general morphology and radial extent of the two images is quite similar we are confident that we have indeed imaged the surface layer of a dusty circumstellar disk. To first order we do not find significant asymmetries in the flux distribution in both filters. Only in the north-western quadrant (roughly at x=0.4$''$ and y=0.4$''$) of the $H$ band image there seems to be a flux excess compared to the corresponding position in the north-eastern quadrant. In the $K_s$ filter image there seems to be a filamentary-like, elongated structure in the northern part of the disk stretching from $\sim$0.5--1.0$''$. However, given that we do not see the same feature in the $H$ filter image and that the total integration time in the $K_s$ image was shorter and hence the signal-to-noise lower, it is unclear whether this feature is real.  

To estimate the disk inclination we fitted ellipses to isophots in the images shown in Figure~\ref{P_images}. While this approach is widely used in the literature is does not take into account potential disk flaring and the scattering (and polarizing) properties of the dust grains. Thus, one has to keep in mind that fitting disk regions of the same brightness does not necessarily mean that one fits the same physical regions. However, this approach gives a first indication about the apparent disk orientation. In the $H$-band image we fitted one ellipse to regions where the surface brightness is between 13 -- 14$\,\rm{mag/arcsec^2}$ and one where it is between  14--15$\,\rm{mag/arcsec^2}$. For the brighter regions the apparent inclination is 34$^\circ\pm$5$^\circ$ from face-on and the position angle of the major axis is 78$^\circ\pm$10$^\circ$ (east of north). For the outer ellipse the inclination is slightly higher with 47$^\circ\pm$2$^\circ$ while the position angle is almost unchanged with 82$^\circ\pm$3$^\circ$. The apparent discrepancy in the disk inclination for the two disk regions might be caused by a combination of a flared disk geometry (see sections 4.2 and 4.3) and the forward/backward scattering and polarization properties of the dust grains. A disk model, which is beyond the scope of the current paper, might help to shed light on this issue. As the $Ks$-band image is not flux calibrated it is not possible to derive a disk inclination that is related to a certain apparent surface brightness level. However, a fit to disk regions that roughly correspond to the same spatial regions as the inner ellipse in the $H$-band yields an inclination and position angle that are identical to those derived for the $H$-band image within the error bars. 
 
Knowing the inclination one can determine the radial brightness profile of the disk along the major axis. Figure~\ref{Radial_profiles} shows the average radial brightness profile computed in a 3 pixels wide box along both semi-major axes. For the orientation of the box we used the average position angle derived from the ellipse fitting exercise described above, i.e., 80$^\circ$. The errors are the root-mean-square of the standard deviation within each box for a given separation and the flux variation observed along both semi-major axes. We then fitted a power-law of the form $S(r)=a\cdot r^x$ to these brightness profiles in both filters between 0.1$''$--1.0$''$ (i.e., $\sim$16 and $\sim$160 AU projected separation). In the $H$ filter the power-law exponent is ${-1.78\pm0.02}$, for the $K_s$ filter the exponent is ${-2.34\pm0.04}$. The fitting functions approximate the data reasonably well. As the surface brightness drops off more steeply in the $K_s$ band, the disk color $[H - K_s]$ in polarized flux becomes bluer at larger separations possibly indicating changes in the dust grain properties. Between 0.6$''$--0.8$''$, however, the fits seem to slightly overestimate the measured profile in both filter. We will discuss this apparent "dip" in more detail in section 4.1.

\section{Discussion}
\subsection{HD97048 in polarized light}
The images presented here are the first resolved NIR scattered light images of the HD97048 disk revealing the inner regions roughly between 16 and 160 AU projected separation. It is interesting to note that the regions we observe here are on the same scale as the orbits of the three outer planets in the HR8799 system \citep[$\sim$24 -- 68 AU;][]{marois2008} in particular because both objects are A-type stars. 

\begin{figure}
\resizebox{\hsize}{!}{\includegraphics{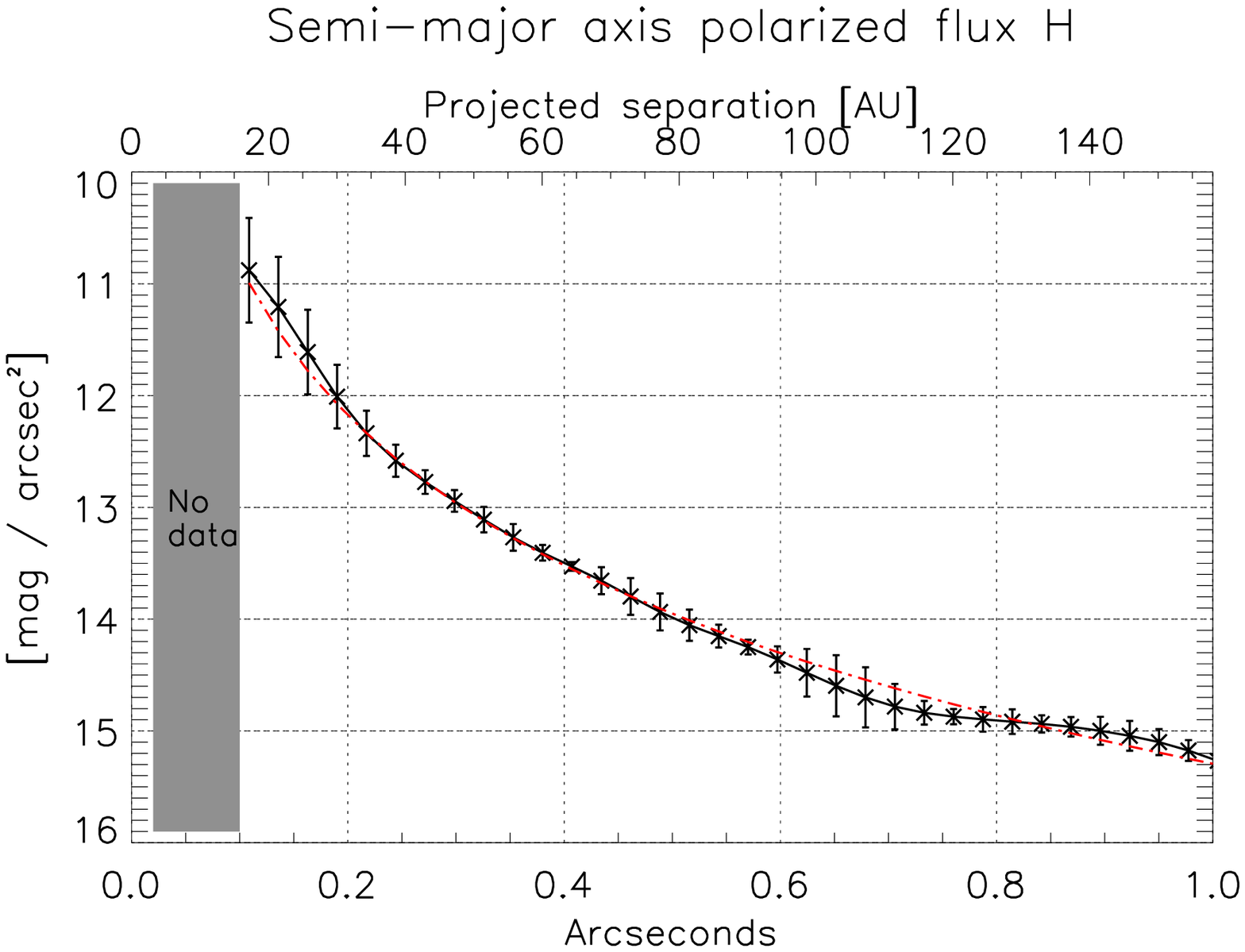}}
\resizebox{\hsize}{!}{\includegraphics{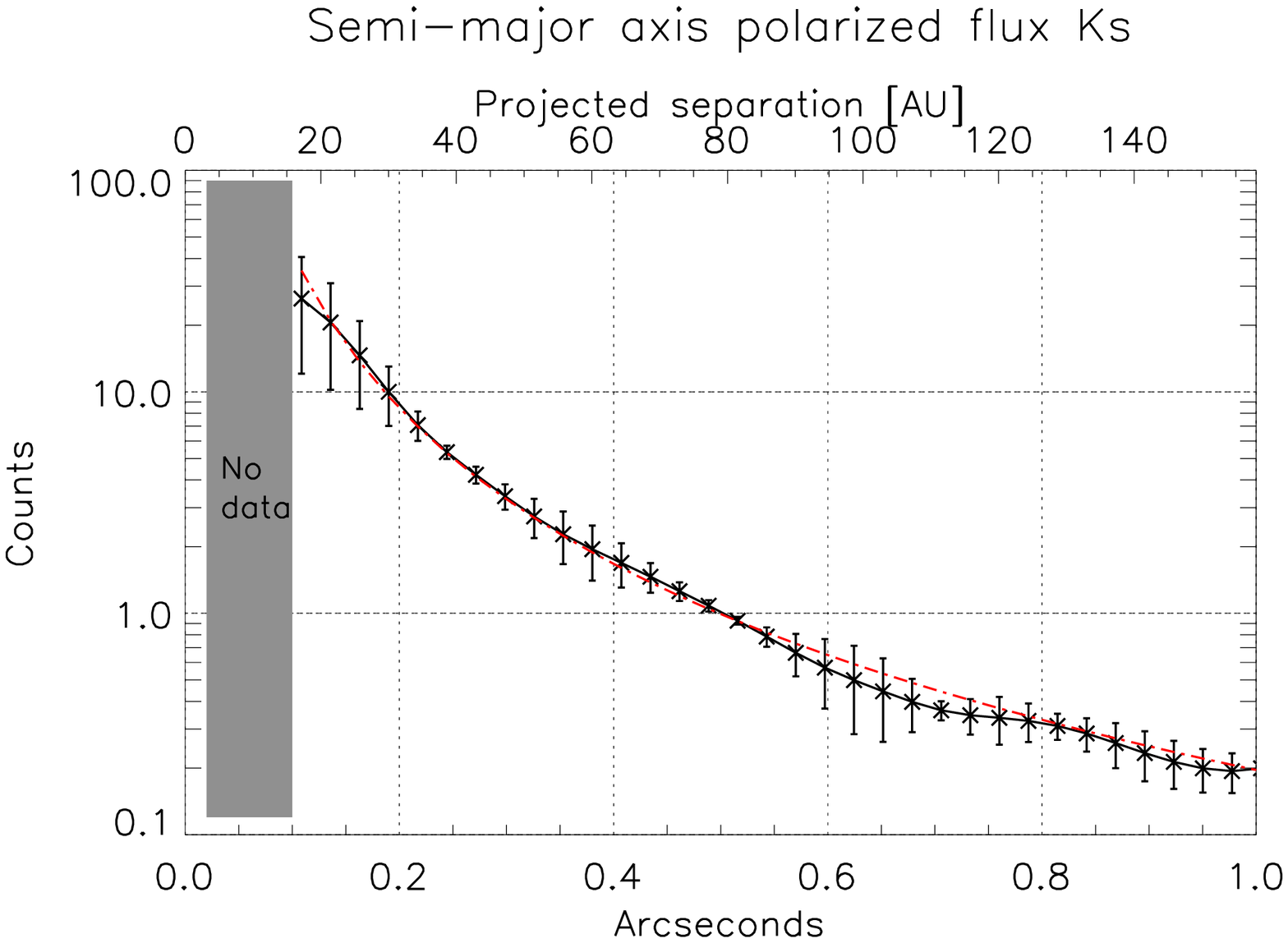}}
\caption{Average surface brightness profiles along both semi-major axes of HD97048} (H filter upper panel, Ks filter lower panel). The data is shown in black data points while the red, dashed lines are power-law fits to the data between 0.1$''$-1.0$''$. The innermost saturated pixels have been masked out, so there is not information available in the inner 0.1$''$.
\label{Radial_profiles}
\end{figure}

As mentioned above the surface brightness profiles of the HD97048 disk contain some indications for a "dip" between 0.6$''$--0.8$''$. In general, such dips in the observed radial brightness profiles could be caused by underlying physical gaps in the disk convolved to the spatial resolution of the observations. In our case, however, this apparent dip is more likely a results of small differences in the flux profiles along both semi-major axes. This also explain the larger error bars of the averaged radial profiles at these separations shown in Figure~\ref{Radial_profiles}.

\subsection{Comparison to HD100546}
It is interesting to compare the results for HD97048 presented here to those for HD100546 presented in \citet{quanz2011}. Not only were both objects observed during the same observing run with the same instrumental setup, but both objects have a similar spectral type and both are members of the Herbig Be/Ae group I in the classification scheme introduced by \citet{meeus2001}. This means that the SED of both objects has a large MIR excess that is generally attributed to arise from a flared disk geometry. In addition, both objects show strong PAH emission bands in the NIR and MIR \citep[e.g.,][and references therein]{vankerckhoven2002,vanboekel2004}. However, there are also notably differences between the two objects. HD100546 shows indications for a large gap in the disk between $\sim$4 -- 15 AU and it was speculated that a gas giant planet orbits in this gap \citep[e.g..][]{bouwman2003,acke2006}. Also, HD97048 lacks a 10 $\mu$m silicate emission feature in the MIR spectrum while HD100546 has a very rich silicate band closely resembling that of comet Hale Bopp \citep{malfait1998,bouwman2003,vanboekel2004}. The lack of a silicate emission feature at 10 $\mu$m on the one hand and the need for a strong opacity source that absorbs effectively in the UV/optical (making the disk flare) and re-emits strongly in the MIR continuum let \citet{vanboekel2004} to the conclusion that small carbonaceous grains are largely present in the HD97048 disk.

Our data allow for the first time a direct comparison of the scattered light properties of the inner disk regions ($\sim$20--140 AU) of HD97048 and HD100546. It shows that the radial brightness profile along the disk major axis is $\propto r^{-3}$ an thus steeper for HD100546 compared to HD97048 where we found roughly $\propto r^{-2}$. In case the emission was optically thin and assuming everything else equal this would indicate that the surface density of scattering dust particles is flatter (i.e., constant) on the surface of the HD97048 disk and/or that dust composition is different. However, in case of an optically thick scattering surface the difference in the radial profiles can be explained with differences in the disk geometry, i.e., stronger disk flaring in case of HD97048. And indeed, \citet{benisty2010} could model the HD100546 disk with a flaring index $\beta=0.5\dots 1.125$ between 13 -- 350 AU, while \citet{lagage2006} modeled the disk around HD 97048 with a flaring index of $\beta=1.26$ (see section 4.3.)\footnote{The flaring geometry of a disk can be described by $H(r)=H_0(r/r_0)^\beta$ with $H(r)$ being the disk surface scale height as a function of the disk radius and $H_0$ and $r_0$ being the scale height and the disk radius at a reference point.}.

\begin{figure}
\resizebox{\hsize}{!}{\includegraphics{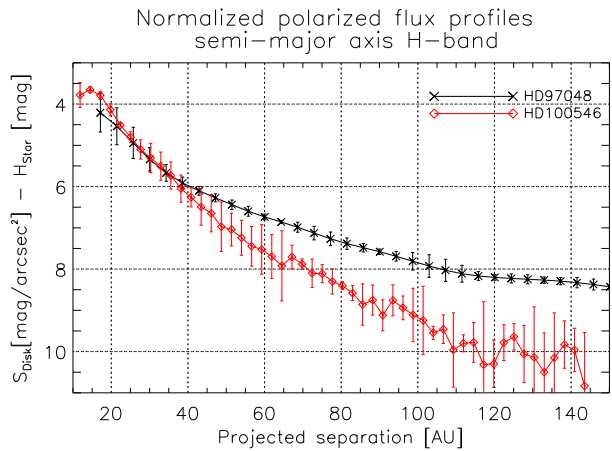}}
\caption{Comparison of the normalized polarized flux profiles of HD97048 (black crosses) and HD100546 \citep[red diamonds;][]{quanz2011} in the $H$ filter. For the normalization the observed $H$-band magnitude of the stars have been subtracted from the surface brightness profiles of the disks. Both curves are the average of the observed surface brightness profiles along both semi-major axes.}
\label{Comparison}
\end{figure}

A direct comparison between the $H$ band surface brightness of the two disks reveals that at 60 AU both disks have roughly the same brightness with 13.5 mag/arcsec$^2$. At smaller projected separations HD100546 appears brighter while at larger separations HD97048 is brighter. Another way of comparing the surface brightness of the two disks is shown in Figure~\ref{Comparison} where we have subtracted the observed $H$-band magnitude of the objects from the disk surface brightness in the $H$ band. The resulting brightness profiles are hence normalized with respect to the total $H$ band flux and provide some information about the general scattering efficiency of the dust. Interestingly, the plot shows that the normalized surface brightness of the the two disks is basically identical in the inner $\sim$20 -- 40 AU. At larger separations the disk of HD97048 appears brighter as the surface brightness profile drops less rapidly (see above). For these analyses one has to keep in mind, though, that the absolute flux calibration for both sources suffers from uncertainties as described in section 2. 

Finally, we can compare the apparent color of the disk as a function of separation. While HD100546 shows in general a red $[H-K_s]$ color that becomes increasingly redder with increasing separation \citep{quanz2011}, HD97048 seems to become relatively bluer at larger separations (see section 3). Concerning the size of the dust grains on the surface of the disk our data provide no constraints for HD97048 as explained in section 3. 

\subsection{HD97048 resolved in PAH, continuum and gas emission}
Since this is the first time the inner disk regions of HD97048 have been directly imaged in scattered light, it's worth putting them into context with other observations probing various disk regions and components.

While earlier mid-infrared observations have shown that the emission of HD97048 is extended on scales up to 5--10$''$ and probably arising from an extended envelope surrounding the star-disk system \citep{prusti1994,siebenmorgen2000}, the first observational evidence that HD97048 might be surrounded by a large, flared circumstellar disk was provided by MIR spectroscopy \citep{vanboekel2004} and later confirmed by direct imaging in PAH filters resolving the disk at 8.6 $\mu$m and 11.3 $\mu$m \citep{lagage2006,doucet2007}.  
\citet{lagage2006} could fit the observations with a flared circumstellar disk with an inclination of $\sim$43$^\circ$ from face-on and flaring geometry described by $H_0\approx 51$ AU, $r_0=135$ AU and $\beta=1.26$ (see footnote 1). While the outer disk radius was found to be at least 370 AU no direct information was obtained for the inner disk regions ($\le$0.5$''$ corresponding to $\lesssim$80 AU) in these MIR images. The disk inclination we find from our scattered light images is roughly in agreement with the value found by \citet{lagage2006} but the position angle of the disk major axis is different. In \citet{lagage2006} and \citet{doucet2007} it seems as if the disk major axis is aligned with the north-south direction giving rise to the observed brightness asymmetry in east-west direction. As seen above, in our images the position angle is close to the east-west direction. We note, however, that we are probing different spatial scales and that, overall, the brightness asymmetry in east-west direction is mostly confined to radial distances $>$0.5$''$ in the PAH filters and not to the MIR continuum \citep{doucet2007, vanboekel2004, marinas2011}.  


\citet{habart2004} showed that emission features at 3.43 and 3.53 $\mu$m, attributed to hydrogenated diamonds \citep[][and references therein]{vankerckhoven2002,habart2004,guillois1999}, as well as the nearby continuum emission was slightly extended and most likely arising from the innermost 15 AU of a circumstellar disk seen face-on. 

\citet{doering2007} used \emph{HST/ACS} to image the circumstellar surrounding of HD97048 in the F606W (broad $V$) filter. They used the 1.8$''$ occulting spot of \emph{HST/ACS} and a reference star for PSF subtraction to reveal scattered light out to a radial distance of $\sim$4$''$ in almost all directions. Since the inner 2$''$ (in radius) were excluded from any analysis due to subtraction residuals these data probe different scales than our observations. \citet{doering2007} found an azimuthally averaged  peak surface brightness of $19.6\pm0.2$ mag/arcsec$^2$ at 2$''$ separation with a radial fall-off $\propto r^{-3.3\pm0.5}$. As this fall-off is significantly steeper than the profiles we find in the inner 1$''$ it seems likely that these data do not represent a simple continuation of our observations to larger radii but that actually different dust populations are probed and it remains unclear whether the \emph{HST/ACS} images can be fully interpreted in the context of an inclined and flared disk or whether the surrounding envelope contributes to the observed properties. 

Looking at the gaseous disk component, \citet{acke2006} found the [OI] emission line broad and double peaked most likely arising from disk regions between 0.8 AU and $\sim$50 AU. \citet{vanderplas2009} resolved several lines in the CO ro-vibrational emission band at 4.7 $\mu$m consistent with being double-peaked and found that the emission extends from 11 AU to at least 100 AU. The lack of emission closer to the star was taken as a possible sign of photo-dissociation of CO as [OI], and hence gas, seems to exits also in the inner 10 AU. Interestingly, \citet{carmona2011} found that the H$_2$ 1-0 S(1) ro-vibrational line at 2.12 $\mu$m was not only clearly detected but extended from $\sim$5-10 AU out to $>$200 AU. 
Photo-dissociation in the inner disk regions and heating by high energy UV and X-ray photons was suggested as a possible origin for the observed emission. Note that also the S(1) pure rotational line of H$_2$ at 17 $\mu$m was detected but unresolved \citep{martinzaidi2007,martinzaidi2009}. 

\begin{figure}
\resizebox{\hsize}{!}{\includegraphics{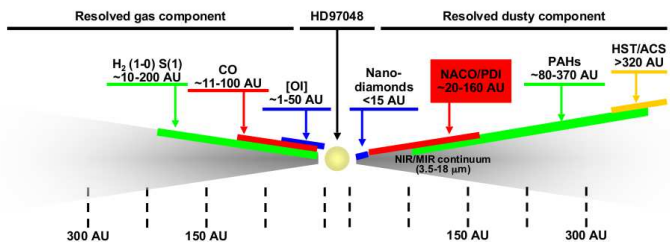}}
\caption{Sketch summarizing the amount of spatially resolved data available for HD97048. The gaseous disk component is shown on the left-hand side, while the dusty disk component is shown on the right-hand side. References for the different observations are given in the text. Our NACO/PDI data are indicated by the red box on the right-hand side.}
\label{sketch}
\end{figure}

It shows that the emitting regions of the gas emission lines are directly comparable to the regions probed by our data. In Figure~\ref{sketch} we have sketched a cartoon of the HD97048 disk showing what sort of spatially resolved information is available. Our NACO/PDI data nicely extend the existing resolved information from the PAH emission closer to the star and probe exactly those regions were several gas species have also been resolved spatially. This makes HD97048 an ideal laboratory for sophisticated disk models studying both disk components - gas and dust - and their interplay with another as well as with the host star.

\section{Summary and conclusions}
For the first time the inner $\sim$0.15$''$--1.0$''$ ($\sim$20--160 AU) of the surface layer of the circumstellar disk around the Herbig Ae/Be star HD97048 have been imaged in scattered light. At least one other early type star is known to host three giant planets at separations comparable to the regions probed by our images \citep[HR8799;][]{marois2008}.
Our data fill a gap in a large collection of existing observations of HD97048 partly resolving the thermal dust emission as well as gas and PAH emission lines on comparable scales. Fitting isophots to the flux calibrated $H$-band images we find an apparent disk inclination of 34$^\circ\pm$5$^\circ$ and 47$^\circ\pm$2$^\circ$ for disk regions with a surface brightness between 13 -- 14$\,\rm{mag/arcsec^2}$ and 14 -- 15$\,\rm{mag/arcsec^2}$, respectively. For the brighter regions the position angle of the major axis is 78$^\circ\pm$10$^\circ$ (east of north) which remains almost unchanged going to the fainter regions (82$^\circ\pm$3$^\circ$).
The surface brightness of the polarized flux drops from $\sim$11 mag/arcsec$^2$ at $\sim$0.1$''$ ($\sim$16 AU) to $\sim$15.3 mag/arcsec$^2$ at $\sim$1.0$''$ ($\sim$160 AU). As the surface brightness drops off more rapidly in $K_s$ compared to $H$ the disks becomes relatively bluer at larger separations possibly indicating changing dust grain properties as a function of radius. The surface brightness profiles along the disk major axis can be fitted with power-laws falling off as $\propto$$r^{-1.78\pm0.02}$ in $H$ and $\propto$$r^{-2.34\pm0.04}$ in $K_s$. In our images, we do not find any significant indications for additional disk structures, e.g., gaps.

The data presented here demonstrate that: 

1) Polarimetric Differential Imaging (PDI) is a very powerful high-contrast technique to image dusty circumstellar disks at inner working angles where we know that planets can form. Other techniques such as classical PSF subtraction or coronagraphy are typically limited to larger separations as the inner disk regions (typically a few tens of AU) either suffer from subtraction residuals or are blocked out by the coronagraph. 

2) Our detection of scattered light around HD97048 makes this object a prime target for future follow-up observations with high-contrast instruments such as SPHERE \citep{beuzit2006} at the VLT or GPI \citep{macintosh2006} at Gemini. These new instruments with their high performance AO systems will provide a much more stable PSF and are thus better suited for high-contrast PDI observations than current instruments. Additionally, SPHERE has also a high-precision imaging polarimeter in the optical wavelength range \citep[ZIMPOL;][]{roelfsema2010} complementing the high contrast capabilities in the NIR. Potentially, SPHERE or GPI will not only confirm our detection but may provide new important insights into the dust grain properties, the disk structure and the interplay between the spatially resolved gas and dust component in the inner regions of the HD97048 circumstellar disk.

\begin{acknowledgements} 
We are very grateful to the people who supported us during the observations, in particular Nancy Ageorges, Markus Hartung, and Nuria Huelamo. SPQ thanks Vincent Geers for useful discussions about PAH emission and Hans Martin Schmid for providing feedback on an earlier version of this manuscript. We acknowledge support from the ESA/ESTEC Faculty Visiting Scientist Programme and thank the anonymous referee for useful comments.
\end{acknowledgements}

\bibliographystyle{aa.bst}
\bibliography{mybib.bib}

\begin{thebibliography}{52}
\expandafter\ifx\csname natexlab\endcsname\relax\def\natexlab#1{#1}\fi

\bibitem[{{Acke} \& {van den Ancker}(2006)}]{acke2006}
{Acke}, B. \& {van den Ancker}, M.~E. 2006, \aap, 449, 267

\bibitem[{{Alibert} {et~al.}(2011){Alibert}, {Mordasini}, \&
  {Benz}}]{alibert2011}
{Alibert}, Y., {Mordasini}, C., \& {Benz}, W. 2011, \aap, 526, A63+

\bibitem[{{Apai} {et~al.}(2008){Apai}, {Janson}, {Moro-Mart{\'{\i}}n}, {Meyer},
  {Mamajek}, {Masciadri}, {Henning}, {Pascucci}, {Kim}, {Hillenbrand},
  {Kasper}, \& {Biller}}]{apai2008}
{Apai}, D., {Janson}, M., {Moro-Mart{\'{\i}}n}, A., {et~al.} 2008, \apj, 672,
  1196

\bibitem[{{Apai} {et~al.}(2004){Apai}, {Pascucci}, {Brandner}, {Henning},
  {Lenzen}, {Potter}, {Lagrange}, \& {Rousset}}]{apai2004}
{Apai}, D., {Pascucci}, I., {Brandner}, W., {et~al.} 2004, \aap, 415, 671

\bibitem[{{Benisty} {et~al.}(2010){Benisty}, {Tatulli}, {M{\'e}nard}, \&
  {Swain}}]{benisty2010}
{Benisty}, M., {Tatulli}, E., {M{\'e}nard}, F., \& {Swain}, M.~R. 2010, \aap,
  511, A75+

\bibitem[{{Beuzit} {et~al.}(2006){Beuzit}, {Feldt}, {Dohlen}, {Mouillet},
  {Puget}, {Antichi}, {Baruffolo}, {Baudoz}, {Berton}, {Boccaletti},
  {Carbillet}, {Charton}, {Claudi}, {Downing}, {Feautrier}, {Fedrigo}, {Fusco},
  {Gratton}, {Hubin}, {Kasper}, {Langlois}, {Moutou}, {Mugnier}, {Pragt},
  {Rabou}, {Saisse}, {Schmid}, {Stadler}, {Turrato}, {Udry}, {Waters}, \&
  {Wildi}}]{beuzit2006}
{Beuzit}, J., {Feldt}, M., {Dohlen}, K., {et~al.} 2006, The Messenger, 125, 29

\bibitem[{{Biller} {et~al.}(2007){Biller}, {Close}, {Masciadri}, {Nielsen},
  {Lenzen}, {Brandner}, {McCarthy}, {Hartung}, {Kellner}, {Mamajek}, {Henning},
  {Miller}, {Kenworthy}, \& {Kulesa}}]{biller2007}
{Biller}, B.~A., {Close}, L.~M., {Masciadri}, E., {et~al.} 2007, \apjs, 173,
  143

\bibitem[{{Borucki} {et~al.}(2011){Borucki}, {Koch}, {Basri}, {Batalha},
  {Brown}, {Bryson}, {Caldwell}, {Christensen-Dalsgaard}, {Cochran}, {DeVore},
  {Dunham}, {Gautier}, {Geary}, {Gilliland}, {Gould}, {Howell}, {Jenkins},
  {Latham}, {Lissauer}, {Marcy}, {Rowe}, {Sasselov}, {Boss}, {Charbonneau},
  {Ciardi}, {Doyle}, {Dupree}, {Ford}, {Fortney}, {Holman}, {Seager},
  {Steffen}, {Tarter}, {Welsh}, {Allen}, {Buchhave}, {Christiansen}, {Clarke},
  {Das}, {D{\'e}sert}, {Endl}, {Fabrycky}, {Fressin}, {Haas}, {Horch},
  {Howard}, {Isaacson}, {Kjeldsen}, {Kolodziejczak}, {Kulesa}, {Li}, {Lucas},
  {Machalek}, {McCarthy}, {MacQueen}, {Meibom}, {Miquel}, {Prsa}, {Quinn},
  {Quintana}, {Ragozzine}, {Sherry}, {Shporer}, {Tenenbaum}, {Torres},
  {Twicken}, {Van Cleve}, {Walkowicz}, {Witteborn}, \& {Still}}]{borucki2011}
{Borucki}, W.~J., {Koch}, D.~G., {Basri}, G., {et~al.} 2011, \apj, 736, 19

\bibitem[{{Bouwman} {et~al.}(2003){Bouwman}, {de Koter}, {Dominik}, \&
  {Waters}}]{bouwman2003}
{Bouwman}, J., {de Koter}, A., {Dominik}, C., \& {Waters}, L.~B.~F.~M. 2003,
  \aap, 401, 577

\bibitem[{{Carmona} {et~al.}(2011){Carmona}, {van der Plas}, {van den Ancker},
  {Audard}, {Waters}, {Fedele}, {Acke}, \& {Pantin}}]{carmona2011}
{Carmona}, A., {van der Plas}, G., {van den Ancker}, M.~E., {et~al.} 2011,
  \aap, 533, A39+

\bibitem[{{Chauvin} {et~al.}(2010){Chauvin}, {Lagrange}, {Bonavita},
  {Zuckerman}, {Dumas}, {Bessell}, {Beuzit}, {Bonnefoy}, {Desidera}, {Farihi},
  {Lowrance}, {Mouillet}, \& {Song}}]{chauvin2010}
{Chauvin}, G., {Lagrange}, A., {Bonavita}, M., {et~al.} 2010, \aap, 509, A52+

\bibitem[{{Cutri} {et~al.}(2003){Cutri}, {Skrutskie}, {van Dyk}, {Beichman},
  {Carpenter}, {Chester}, {Cambresy}, {Evans}, {Fowler}, {Gizis}, {Howard},
  {Huchra}, {Jarrett}, {Kopan}, {Kirkpatrick}, {Light}, {Marsh}, {McCallon},
  {Schneider}, {Stiening}, {Sykes}, {Weinberg}, {Wheaton}, {Wheelock}, \&
  {Zacarias}}]{cutri2003}
{Cutri}, R.~M., {Skrutskie}, M.~F., {van Dyk}, S., {et~al.} 2003, {2MASS All
  Sky Catalog of point sources.} (The IRSA 2MASS All-Sky Point Source Catalog,
  NASA/IPAC Infrared Science
  Archive.~http://irsa.ipac.caltech.edu/applications/Gator/)

\bibitem[{{Doering} {et~al.}(2007){Doering}, {Meixner}, {Holfeltz}, {Krist},
  {Ardila}, {Kamp}, {Clampin}, \& {Lubow}}]{doering2007}
{Doering}, R.~L., {Meixner}, M., {Holfeltz}, S.~T., {et~al.} 2007, \aj, 133,
  2122

\bibitem[{{Doucet} {et~al.}(2007){Doucet}, {Habart}, {Pantin}, {Dullemond},
  {Lagage}, {Pinte}, {Duch{\^e}ne}, \& {M{\'e}nard}}]{doucet2007}
{Doucet}, C., {Habart}, E., {Pantin}, E., {et~al.} 2007, \aap, 470, 625

\bibitem[{{Guillois} {et~al.}(1999){Guillois}, {Ledoux}, \&
  {Reynaud}}]{guillois1999}
{Guillois}, O., {Ledoux}, G., \& {Reynaud}, C. 1999, \apjl, 521, L133

\bibitem[{{Habart} {et~al.}(2004){Habart}, {Testi}, {Natta}, \&
  {Carbillet}}]{habart2004}
{Habart}, E., {Testi}, L., {Natta}, A., \& {Carbillet}, M. 2004, \apjl, 614,
  L129

\bibitem[{{Henning} {et~al.}(1998){Henning}, {Burkert}, {Launhardt}, {Leinert},
  \& {Stecklum}}]{henning1998}
{Henning}, T., {Burkert}, A., {Launhardt}, R., {Leinert}, C., \& {Stecklum}, B.
  1998, \aap, 336, 565

\bibitem[{{Janson} {et~al.}(2011){Janson}, {Bonavita}, {Klahr},
  {Lafreni{\`e}re}, {Jayawardhana}, \& {Zinnecker}}]{janson2011}
{Janson}, M., {Bonavita}, M., {Klahr}, H., {et~al.} 2011, \apj, 736, 89

\bibitem[{{Kasper} {et~al.}(2007){Kasper}, {Apai}, {Janson}, \&
  {Brandner}}]{kasper2007}
{Kasper}, M., {Apai}, D., {Janson}, M., \& {Brandner}, W. 2007, \aap, 472, 321

\bibitem[{{Kuhn} {et~al.}(2001){Kuhn}, {Potter}, \& {Parise}}]{kuhn2001}
{Kuhn}, J.~R., {Potter}, D., \& {Parise}, B. 2001, \apjl, 553, L189

\bibitem[{{Lafreni{\`e}re} {et~al.}(2007){Lafreni{\`e}re}, {Doyon}, {Marois},
  {Nadeau}, {Oppenheimer}, {Roche}, {Rigaut}, {Graham}, {Jayawardhana},
  {Johnstone}, {Kalas}, {Macintosh}, \& {Racine}}]{lafreniere2007}
{Lafreni{\`e}re}, D., {Doyon}, R., {Marois}, C., {et~al.} 2007, \apj, 670, 1367

\bibitem[{{Lagage} {et~al.}(2006){Lagage}, {Doucet}, {Pantin}, {Habart},
  {Duch{\^e}ne}, {M{\'e}nard}, {Pinte}, {Charnoz}, \& {Pel}}]{lagage2006}
{Lagage}, P.-O., {Doucet}, C., {Pantin}, E., {et~al.} 2006, Science, 314, 621

\bibitem[{{Lagrange} {et~al.}(2009){Lagrange}, {Gratadour}, {Chauvin}, {Fusco},
  {Ehrenreich}, {Mouillet}, {Rousset}, {Rouan}, {Allard}, {Gendron}, {Charton},
  {Mugnier}, {Rabou}, {Montri}, \& {Lacombe}}]{lagrange2009a}
{Lagrange}, A., {Gratadour}, D., {Chauvin}, G., {et~al.} 2009, \aap, 493, L21

\bibitem[{Lagrange {et~al.}(2010)Lagrange, Bonnefoy, Chauvin, Apai, Ehrenreich,
  Boccaletti, Gratadour, Rouan, Mouillet, Lacour, \& Kasper}]{lagrange2010}
Lagrange, A.-M., Bonnefoy, M., Chauvin, G., {et~al.} 2010, Science,
  science.1187187

\bibitem[{{Lenzen} {et~al.}(2003){Lenzen}, {Hartung}, {Brandner}, {Finger},
  {Hubin}, {Lacombe}, {Lagrange}, {Lehnert}, {Moorwood}, \&
  {Mouillet}}]{lenzen2003}
{Lenzen}, R., {Hartung}, M., {Brandner}, W., {et~al.} 2003, in Society of
  Photo-Optical Instrumentation Engineers (SPIE) Conference Series, Vol. 4841,
  Society of Photo-Optical Instrumentation Engineers (SPIE) Conference Series,
  ed. {M.~Iye \& A.~F.~M.~Moorwood}, 944--952

\bibitem[{{Macintosh} {et~al.}(2006){Macintosh}, {Graham}, {Palmer}, {Doyon},
  {Gavel}, {Larkin}, {Oppenheimer}, {Saddlemyer}, {Wallace}, {Bauman}, {Evans},
  {Erikson}, {Morzinski}, {Phillion}, {Poyneer}, {Sivaramakrishnan}, {Soummer},
  {Thibault}, \& {Veran}}]{macintosh2006}
{Macintosh}, B., {Graham}, J., {Palmer}, D., {et~al.} 2006, in Society of
  Photo-Optical Instrumentation Engineers (SPIE) Conference Series, Vol. 6272,
  Society of Photo-Optical Instrumentation Engineers (SPIE) Conference Series

\bibitem[{{Malfait} {et~al.}(1998){Malfait}, {Waelkens}, {Waters},
  {Vandenbussche}, {Huygen}, \& {de Graauw}}]{malfait1998}
{Malfait}, K., {Waelkens}, C., {Waters}, L.~B.~F.~M., {et~al.} 1998, \aap, 332,
  L25

\bibitem[{{Mari{\~n}as} {et~al.}(2011){Mari{\~n}as}, {Telesco}, {Fisher}, \&
  {Packham}}]{marinas2011}
{Mari{\~n}as}, N., {Telesco}, C.~M., {Fisher}, R.~S., \& {Packham}, C. 2011,
  \apj, 737, 57

\bibitem[{{Marois} {et~al.}(2008){Marois}, {Macintosh}, {Barman}, {Zuckerman},
  {Song}, {Patience}, {Lafreni{\`e}re}, \& {Doyon}}]{marois2008}
{Marois}, C., {Macintosh}, B., {Barman}, T., {et~al.} 2008, Science, 322, 1348

\bibitem[{{Marois} {et~al.}(2010){Marois}, {Zuckerman}, {Konopacky},
  {Macintosh}, \& {Barman}}]{marois2010}
{Marois}, C., {Zuckerman}, B., {Konopacky}, Q.~M., {Macintosh}, B., \&
  {Barman}, T. 2010, ArXiv e-prints

\bibitem[{{Martin-Za{\"i}di} {et~al.}(2009){Martin-Za{\"i}di}, {Habart},
  {Augereau}, {M{\'e}nard}, {Lagage}, {Pantin}, \&
  {Olofsson}}]{martinzaidi2009}
{Martin-Za{\"i}di}, C., {Habart}, E., {Augereau}, J.-C., {et~al.} 2009, \apj,
  695, 1302

\bibitem[{{Martin-Za{\"i}di} {et~al.}(2007){Martin-Za{\"i}di}, {Lagage},
  {Pantin}, \& {Habart}}]{martinzaidi2007}
{Martin-Za{\"i}di}, C., {Lagage}, P.-O., {Pantin}, E., \& {Habart}, E. 2007,
  \apjl, 666, L117

\bibitem[{{Masciadri} {et~al.}(2005){Masciadri}, {Mundt}, {Henning}, {Alvarez},
  \& {Barrado y Navascu{\'e}s}}]{masciadri2005}
{Masciadri}, E., {Mundt}, R., {Henning}, T., {Alvarez}, C., \& {Barrado y
  Navascu{\'e}s}, D. 2005, \apj, 625, 1004

\bibitem[{{Mayor} {et~al.}(2011){Mayor}, {Marmier}, {Lovis}, {Udry},
  {S{\'e}gransan}, {Pepe}, {Benz}, {Bertaux}, {Bouchy}, {Dumusque}, {Lo Curto},
  {Mordasini}, {Queloz}, \& {Santos}}]{mayor2011}
{Mayor}, M., {Marmier}, M., {Lovis}, C., {et~al.} 2011, ArXiv e-prints

\bibitem[{{Meeus} {et~al.}(2001){Meeus}, {Waters}, {Bouwman}, {van den Ancker},
  {Waelkens}, \& {Malfait}}]{meeus2001}
{Meeus}, G., {Waters}, L.~B.~F.~M., {Bouwman}, J., {et~al.} 2001, \aap, 365,
  476

\bibitem[{{Perrin} {et~al.}(2004){Perrin}, {Graham}, {Kalas}, {Lloyd}, {Max},
  {Gavel}, {Pennington}, \& {Gates}}]{perrin2004}
{Perrin}, M.~D., {Graham}, J.~R., {Kalas}, P., {et~al.} 2004, Science, 303,
  1345

\bibitem[{{Potter}(2003)}]{potter2003}
{Potter}, D.~E. 2003, PhD thesis, UNIVERSITY OF HAWAI'I

\bibitem[{{Prusti} {et~al.}(1994){Prusti}, {Natta}, \& {Palla}}]{prusti1994}
{Prusti}, T., {Natta}, A., \& {Palla}, F. 1994, \aap, 292, 593

\bibitem[{{Quanz} {et~al.}(2010){Quanz}, {Meyer}, {Kenworthy}, {Girard},
  {Kasper}, {Lagrange}, {Apai}, {Boccaletti}, {Bonnefoy}, {Chauvin}, {Hinz}, \&
  {Lenzen}}]{quanz2010}
{Quanz}, S.~P., {Meyer}, M.~R., {Kenworthy}, M.~A., {et~al.} 2010, \apjl, 722,
  L49

\bibitem[{{Quanz} {et~al.}(2011){Quanz}, {Schmid}, {Geissler}, {Meyer},
  {Henning}, {Brandner}, \& {Wolf}}]{quanz2011}
{Quanz}, S.~P., {Schmid}, H.~M., {Geissler}, K., {et~al.} 2011, \apj, 738, 23

\bibitem[{{Roelfsema} {et~al.}(2010){Roelfsema}, {Schmid}, {Pragt}, {Gisler},
  {Waters}, {Bazzon}, {Baruffolo}, {Beuzit}, {Boccaletti}, {Charton}, {Cumani},
  {Dohlen}, {Downing}, {Elswijk}, {Feldt}, {Groothuis}, {de Haan}, {Hanenburg},
  {Hubin}, {Joos}, {Kasper}, {Keller}, {Kragt}, {Lizon}, {Mouillet}, {Pavlov},
  {Rigal}, {Rochat}, {Salasnich}, {Steiner}, {Thalmann}, {Venema}, \&
  {Wildi}}]{roelfsema2010}
{Roelfsema}, R., {Schmid}, H.~M., {Pragt}, J., {et~al.} 2010, in Society of
  Photo-Optical Instrumentation Engineers (SPIE) Conference Series, Vol. 7735,
  Society of Photo-Optical Instrumentation Engineers (SPIE) Conference Series

\bibitem[{{Rousset} {et~al.}(2003){Rousset}, {Lacombe}, {Puget}, {Hubin},
  {Gendron}, {Fusco}, {Arsenault}, {Charton}, {Feautrier}, {Gigan}, {Kern},
  {Lagrange}, {Madec}, {Mouillet}, {Rabaud}, {Rabou}, {Stadler}, \&
  {Zins}}]{rousset2003}
{Rousset}, G., {Lacombe}, F., {Puget}, P., {et~al.} 2003, in Society of
  Photo-Optical Instrumentation Engineers (SPIE) Conference Series, Vol. 4839,
  Society of Photo-Optical Instrumentation Engineers (SPIE) Conference Series,
  ed. {P.~L.~Wizinowich \& D.~Bonaccini}, 140--149

\bibitem[{{Siebenmorgen} {et~al.}(2000){Siebenmorgen}, {Prusti}, {Natta}, \&
  {M{\"u}ller}}]{siebenmorgen2000}
{Siebenmorgen}, R., {Prusti}, T., {Natta}, A., \& {M{\"u}ller}, T.~G. 2000,
  \aap, 361, 258

\bibitem[{{van Boekel} {et~al.}(2004){van Boekel}, {Waters}, {Dominik},
  {Dullemond}, {Tielens}, \& {de Koter}}]{vanboekel2004}
{van Boekel}, R., {Waters}, L.~B.~F.~M., {Dominik}, C., {et~al.} 2004, \aap,
  418, 177

\bibitem[{{van den Ancker} {et~al.}(1998){van den Ancker}, {de Winter}, \&
  {Tjin A Djie}}]{vandenancker1998}
{van den Ancker}, M.~E., {de Winter}, D., \& {Tjin A Djie}, H.~R.~E. 1998,
  \aap, 330, 145

\bibitem[{{van den Ancker} {et~al.}(1997){van den Ancker}, {The}, {Tjin A
  Djie}, {Catala}, {de Winter}, {Blondel}, \& {Waters}}]{vandenancker1997}
{van den Ancker}, M.~E., {The}, P.~S., {Tjin A Djie}, H.~R.~E., {et~al.} 1997,
  \aap, 324, L33

\bibitem[{{van der Plas} {et~al.}(2009){van der Plas}, {van den Ancker},
  {Acke}, {Carmona}, {Dominik}, {Fedele}, \& {Waters}}]{vanderplas2009}
{van der Plas}, G., {van den Ancker}, M.~E., {Acke}, B., {et~al.} 2009, \aap,
  500, 1137

\bibitem[{{Van Kerckhoven} {et~al.}(2002){Van Kerckhoven}, {Tielens}, \&
  {Waelkens}}]{vankerckhoven2002}
{Van Kerckhoven}, C., {Tielens}, A.~G.~G.~M., \& {Waelkens}, C. 2002, \aap,
  384, 568

\bibitem[{{van Leeuwen}(2007)}]{vanleeuwen2007}
{van Leeuwen}, F. 2007, \aap, 474, 653

\bibitem[{{Whittet} {et~al.}(1987){Whittet}, {Kirrane}, {Kilkenny}, {Oates},
  {Watson}, \& {King}}]{whittet1987}
{Whittet}, D.~C.~B., {Kirrane}, T.~M., {Kilkenny}, D., {et~al.} 1987, \mnras,
  224, 497

\bibitem[{{Whittet} {et~al.}(1997){Whittet}, {Prusti}, {Franco}, {Gerakines},
  {Kilkenny}, {Larson}, \& {Wesselius}}]{whittet1997}
{Whittet}, D.~C.~B., {Prusti}, T., {Franco}, G.~A.~P., {et~al.} 1997, \aap,
  327, 1194

\bibitem[{{Witzel} {et~al.}(2011){Witzel}, {Eckart}, {Buchholz}, {Zamaninasab},
  {Lenzen}, {Sch{\"o}del}, {Araujo}, {Sabha}, {Bremer}, {Karas}, {Straubmeier},
  \& {Muzic}}]{witzel2011}
{Witzel}, G., {Eckart}, A., {Buchholz}, R.~M., {et~al.} 2011, \aap, 525, A130+

\end{thebibliography}

\end{document}